\documentclass[groupedaddress,showpacs,a4paper,showkeys]{revtex4}
\usepackage{amsmath,amsthm,amssymb}
\newcommand{\al}{\alpha}
\newcommand{\be}{\beta}
\newcommand{\ep}{\epsilon}
\newcommand{\ga}{\gamma}
\newcommand{\de}{\delta}
\newcommand{\la}{\lambda}
\newcommand{\om}{\omega}
\newcommand{\si}{\sigma}
\newcommand{\te}{\theta}
\newcommand{\vp}{\varphi}
\newcommand{\ze}{\zeta}
\newcommand{\De}{\Delta}
\newcommand{\La}{\Lambda}
\newcommand{\bx}{\mathbf{x}}
\newcommand{\bsi}{\boldsymbol{\si}}
\newcommand{\bw}{\mathbf{w}}
\newcommand{\bz}{\mathbf{z}}
\newcommand{\bze}{\boldsymbol{\ze}}
\newcommand{\tc}{\tilde{c}}
\newcommand{\tl}{\tilde{l}}
\newcommand{\tH}{\widetilde{H}}
\newcommand{\tK}{\tilde{K}}
\newcommand{\tS}{\tilde{S}}
\newcommand{\NN}{{\mathbb N}}
\newcommand{\RR}{{\mathbb R}}
\newcommand{\ZZ}{{\mathbb Z}}
\newcommand{\cM}{{\mathcal M}}
\newcommand{\cN}{{\mathcal N}}
\newcommand{\cP}{{\mathcal P}}
\newcommand{\cR}{{\mathcal R}}

\newcommand{\fsl}{\mathfrak{sl}}
\newcommand{\fS}{{\mathfrak S}}
\def\BC{\,\overline{\!C}{}}
\def\BH{\,\overline{\!H}{}}
\def\BcP{\overline{\cP}{}}
\def\BcR{\overline{\cR}{}}
\newcommand{\pa}{\partial}
\newcommand{\ra}{\rightarrow}

\newcommand{\nid}{\noindent}
\newcommand{\abs}[1]{\left|#1\right|}
\def\ket#1#2{|{#1}{#2}\rangle}
\def\kett#1#2#3{|{#1}{#2}{#3}\rangle}
\def\Strut{\vrule height12pt width0pt}
\def\BStrut{\vrule height16pt depth10pt width0pt}
\let\ds\displaystyle
\newcommand{\sign}{\operatorname{sign}}
\newcommand{\myspan}{\operatorname{span}}
\newcommand{\diff}{\operatorname{d}\!}
\newcommand{\sn}{\operatorname{sn}}
\newcommand{\cn}{\operatorname{cn}}
\newcommand{\dn}{\operatorname{dn}}
\newcommand{\iu}{{\rm i}}
\newcommand{\e}{{\rm e}}
\begin{document}
\title{New spin Calogero--Sutherland models related to $B_N$-type
Dunkl operators}
\author{F. Finkel}
\author{D. G\'omez-Ullate}
\author{A. Gonz\'alez-L\'opez}
\email[Corresponding author. E-mail: ]{artemio@eucmos.sim.ucm.es}
\author{M. A. Rodr\'{\i}guez}
\affiliation{Departamento de F\'{\i}sica Te\'orica II, Universidad
Complutense, 28040 Madrid, Spain}
\author{R. Zhdanov}
\altaffiliation{On leave of absence from Institute of
Mathematics, 3 Tereshchenkivska St., 01601 Kyiv - 4 Ukraine}
\affiliation{Departamento de F\'{\i}sica Te\'orica II, Universidad
Complutense, 28040 Madrid, Spain}
\date{April 18, 2001}

\begin{abstract}\vskip 2mm
We construct several new families of exactly and quasi-exactly solvable
$BC_N$-type Calogero--Sutherland models with internal degrees of freedom.
Our approach is based on the introduction of a new family of
Dunkl operators of $B_N$ type which, together with the
original $B_N$-type Dunkl operators, are shown to preserve certain polynomial
subspaces of finite dimension. We prove that a wide class of quadratic
combinations involving these three sets of Dunkl operators always yields
a spin Calogero--Sutherland model, which is (quasi-)exactly solvable by
construction. We show that all the spin Calogero--Sutherland models
obtainable within this framework can be expressed in a unified way in
terms of a Weierstrass $\wp$ function with suitable half-periods.
This provides a natural spin counterpart of the well-known general formula
for a scalar completely integrable potential of $BC_N$ type due to
Olshanetsky and Perelomov. As an illustration of our method, we exactly compute
several energy levels and their corresponding wavefunctions of
an elliptic quasi-exactly solvable potential for two and
three particles of spin $1/2$.
\end{abstract}

\maketitle
\section{Introduction}

The completely integrable and exactly solvable models of
Calogero~\cite{Ca71} and Sutherland~\cite{Su7172} describe a
system of $N$ quantum particles in one dimension with long-range
pairwise interaction. These models and their subsequent
generalizations (see~\cite{OP83} and references therein for a
comprehensive review) have been extensively applied in many
different fields of physical interest, such as fractional
statistics and anyons~\cite{Po89,Ha9495,ILO99}, quantum Hall
liquids~\cite{AI94}, Yang--Mills theories~\cite{GN94,DF98}, and
propagation of soliton waves~\cite{Po95}. A significant effort
has been devoted over the last decade to the extension
of scalar Calogero--Sutherland models to systems of particles with
internal degrees of freedom or
``spin''~\cite{HH92,BGHP93,HW93,MP93,MP94,Ya95,BM96,In97,Du98,FGGRZ01}.
These models have attracted considerable
interest due to their connection with integrable spin chains of Haldane--Shastry
type~\cite{Ha88,Sh88} through the ``freezing trick'' of
Polychronakos~\cite{Po93}.

The exactly solvable and integrable spin models
introduced in~\cite{HH92,MP93} generalize
the original rational (Calogero) and trigonometric
(Sutherland) scalar models, and are
invariant with respect to the
Weyl group of type $A_N$.
The exact solvability of both models
can be established by relating the Hamiltonian to a
quadratic combination of either the Dunkl~\cite{Du89}
or the Dunkl--Cherednik~\cite{Ch9194} operators of $A_N$ type,
whose relevance in this context was first pointed out by
Polychronakos~\cite{Po92}.
We shall use the term ``Dunkl operators'' to collectively refer to
this type of operators.
Up to the best of our knowledge, only two $B_N$-invariant
spin Calogero--Sutherland models have been proposed so far
in the literature, namely the rational and the
trigonometric spin models constructed by Yamamoto in~\cite{Ya95}.
The exact solvability of the rational Yamamoto model was later
proved in Ref.~\cite{Du98} using the Dunkl operator formalism.
The exact solvability of the
trigonometric Yamamoto model will be proved in this paper.

In a recent paper~\cite{FGGRZ01} the authors proposed a new
systematic method for constructing spin Calogero--Sutherland
models of type $A_N$. One of the key ingredients of the
method was the introduction of a new family of Dunkl-type operators which,
together with the Dunkl operators defined in~\cite{Du89,Ch9194},
preserve a certain polynomial module of finite dimension.
It was shown that a wide class of quadratic combinations of all three
types of Dunkl operators always yields a spin Calogero--Sutherland
model. In this way
all the previously known exactly solvable spin Calogero--Sutherland
models of $A_N$ type are recovered and, what is more important,
several new exactly and {\em quasi-exactly} solvable spin models are obtained.
By quasi-exactly solvable (QES) we mean here that
the Hamiltonian preserves a known finite-dimensional
subspace of smooth functions, so that a finite
subset of the spectrum can be computed algebraically;
see~\cite{Tu88,Sh89,Us94} for further details.
If the Hamiltonian leaves invariant an infinite increasing
sequence of finite-dimensional subspaces, we shall say that the
model is exactly solvable (ES).

In this paper we extend the method of Ref.~\cite{FGGRZ01} to construct
new families of (Q)ES spin Calogero--Sutherland models of $BC_N$ type.
To this end, we define in Section~\ref{sec.dunkl}
a new set of Dunkl operators of $B_N$ type
leaving invariant a certain polynomial subspace of finite dimension,
which is also preserved by the
original Dunkl operators of $B_N$ type introduced in~\cite{Du98}.
In Section~\ref{sec.ham}, we show that a suitable
quadratic combination of all three types of Dunkl operators
discussed in Section~\ref{sec.dunkl} can be mapped into a
multi-parameter (Q)ES physical Hamiltonian with spin.
This approach is a generalization of the construction used
to prove the integrability of the $A_N$ spin Calogero--Sutherland
models, in which only a {\em single} set of Dunkl operators
is involved. Our method is also related to the so-called
{\em hidden symmetry algebra}
approach to scalar $N$-body QES models~\cite{MRT96,HS99,GGR00},
where the Hamiltonian is expressed as a quadratic combination
of the generators of a realization of $\fsl(N+1)$.
We then show that the sets of Dunkl operators used in our construction
are invariant under inversions and scale transformations. This
property is exploited in Section~\ref{sec.class} to perform
a complete classification of the $BC_N$-type (Q)ES spin Calogero--Sutherland
models that can be constructed with the method described in this paper.
The resulting potentials can be divided into nine inequivalent classes,
out of which only two (the rational and trigonometric Yamamoto
models) were previously known. In particular, we obtain
four new families of elliptic
QES spin Calogero--Sutherland models of $BC_N$ type.
Section~\ref{sec.disc} is devoted to the discussion of the
general structure of the potentials listed in Section~\ref{sec.class}.
We prove that all the potentials in the classification are expressible
in a unified way in terms of a Weierstrass $\wp$
function with suitable (sometimes infinite) half-periods.
This provides a natural spin counterpart of Olshanetsky and
Perelomov's formula for a general scalar potential
related to the $BC_N$ root system. Finally, in Section~\ref{sec.ex}
we illustrate the method by exactly computing several energy
levels and their corresponding eigenstates for an elliptic
spin $1/2$ potential in the two- and three-particle cases.

\section{$B_N$-type Dunkl Operators}\label{sec.dunkl}

In this section we introduce a new familiy of $B_N$-type Dunkl operators
which will play a central role in our construction of new (Q)ES spin
Calogero--Sutherland models.

Let $f(\bz)$ be an arbitrary function of
$\bz=(z_1,\dots,z_N)\in\RR^N$. Consider the permutation operators
$K_{ij}=K_{ji}$ and the sign reversing operators $K_i$, whose
action on the function $f$ is given by
\begin{equation}
\begin{aligned}
& (K_{ij}f)(z_1,\dots,z_i,\dots,z_j,\dots,z_N)=f(z_1,\dots,
z_j,\dots,z_i,\dots,z_N)\,,\\
& (K_i f)(z_1,\dots,z_i,\dots,z_N)=f(z_1,\dots,-z_i,\dots,z_N)\,,
\label{KR}
\end{aligned}
\end{equation}
where $i,j=1,\dots,N$. It follows that $K_{ij}$ and $K_i$ verify
the relations
\begin{equation}
\begin{gathered}
    K_{ij}^2=1,\qquad K_{ij}K_{jk}=K_{ik}K_{ij}=K_{jk}K_{ik},\qquad
    K_{ij}K_{kl}=K_{kl}K_{ij},\\
    K_i^2=1,\qquad K_iK_j=K_jK_i\,,\qquad K_{ij}K_k=K_k K_{ij},\qquad
    K_{ij}K_j=K_i K_{ij}\,,
\end{gathered}
    \label{KRalg}
\end{equation}
    where the indices $i,j,k,l$ take distinct values in the range
$1,\dots,N$. The operators $K_{ij}, K_i$ span the Weyl group of
type $B_N$, also called the hyperoctahedral group. We shall also
employ the customary notation $\tK_{ij}=K_iK_jK_{ij}$. Let us
consider the following set of Dunkl operators:
\begin{align}
J_i^- &=\frac\pa{\pa z_i}+a\Bigg(\sum_{j\neq
i}\frac1{z_i-z_j}\,(1-K_{ij})+\sum_{j\neq
i}\frac1{z_i+z_j}\,(1-\tK_{ij})\Bigg)+\frac b{z_i}\,(1-K_i)\,,\label{J-}\\
J_i^0 &= z_i\frac\pa{\pa z_i}-\frac m2+\frac a2 \Bigg(\sum_{j\neq
i}\frac{z_i+z_j}{z_i-z_j}\,(1-K_{ij})+\sum_{j\neq
i}\frac{z_i-z_j}{z_i+z_j}\,(1-\tK_{ij})\Bigg),\label{J0}\\
J_i^+ &=z_i^2\frac\pa{\pa z_i}-m\,z_i+a\Bigg(\sum_{j\neq
i}\frac{z_iz_j}{z_i-z_j}\,(1-K_{ij})-\sum_{j\neq
i}\frac{z_iz_j}{z_i+z_j}\,(1-\tK_{ij})\Bigg)-b'z_i
\big(1-(-1)^m K_i\big),\label{J+}
\end{align}
where $a,b,b'$ are nonzero real parameters, $m$ is a nonnegative integer, and
$i=1,\dots,N$. In Eqs.~\eqref{J-}--\eqref{J+}, the symbol
$\sum_{j\neq i}$ denotes summation in $j$ with
$j=1,\dots,i-1,i+1,\dots,N$. In general, any summation or product index
without an explicit range will be understood in this paper
to run from $1$ to $N$, unless otherwise constrained.
It shall also be clear in each case whether a sum symbol
with more than one index present denotes single or multiple summation.

The operators $J_i^-$ in Eq.~\eqref{J-} have been used by
Dunkl~\cite{Du98} to construct a complete set of eigenvectors for
Yamamoto's $B_N$ rational spin model~\cite{Ya95}.
The operators $J_i^0$ were also introduced by Dunkl in Ref.~\cite{Du98}.
To the best of our knowledge, the operators
$J_i^+$ have not been considered previously in the literature.

The operators~\eqref{J-}--\eqref{J+} obey the commutation
relations
\begin{gather}
[J_i^{\pm},J_j^{\pm}]=0\,,\qquad
[J_i^0,J_j^0]=\frac{a^2}4\sum_{k\neq i,j}(K_{ij}+\tK_{ij})
(K_{jk}+\tK_{jk}-K_{ik}-\tK_{ik})\,,\label{commJ}\\
[K_{ij},J_k^\ep]=0\,,\qquad K_{ij}J_i^\ep=J_j^\ep K_{ij}\,,\qquad
[K_i,J_j^\ep]=0\,,\qquad K_iJ_i^\ep=(-1)^\ep J_i^\ep K_i\,,\label{commJKR}
\end{gather}
where $\ep=\pm,0$, and the indices $i,j,k$ take distinct values in the
range $1,\dots,N$. The operators $J_i^-$ (respectively $J_i^+$),
$i=1,\dots,N$, together with $K_{ij}$ and $K_i$, $i,j=1,\dots,N$, span
a degenerate affine Hecke algebra, see~\cite{Ch9194}. The operators
$J_i^0$ do not commute, but since $J_i^0+\frac
a2\sum_{j<i}(K_{ij}+\tK_{ij}) -\frac a2\sum_{j>i}(K_{ij}+\tK_{ij})$
do, it can be shown that the latter operators, together with $K_{ij}$
and $K_i$, also define a degenerate affine Hecke algebra.

It is well-known~\cite{Du98} that the operators $J_i^-$
and $J_i^0$ preserve the space
$\cP_n$ of polynomials in $z_1,\dots,z_N$ of degree at most $n$,
for all $n\in\NN$. Moreover, for {\em any} nonnegative integer $n$,
the space $\cR_n$ spanned by the monomials
$\prod_i z_i^{l_i}$ with $0\leq l_i\leq n$ is also invariant
under the action of both $J_i^-$ and $J_i^0$.
Let us prove this assertion in the case of $J_i^0$. Since
$(z_i\frac\pa{\pa z_i}-\frac m2)\cR_n\subset\cR_n$,
it suffices to show that
\begin{equation}
    \frac{z_i+z_j}{z_i-z_j}\,(1-K_{ij})\prod_k z_k^{l_k}\subset\cR_n\,,
    \qquad\text{and}\qquad
    \frac{z_i-z_j}{z_i+z_j}\,(1-\tK_{ij})\prod_k z_k^{l_k}\subset\cR_n\,,
    \label{inclusions}
\end{equation}
for any pair of indices $1\leq i\neq j\leq N$. For the first inclusion
we note that
\begin{equation}
\begin{aligned}
\frac{z_i+z_j}{z_i-z_j}\,(1-K_{ij})\prod_k z_k^{l_k}
&=\Bigg(\prod_{k\neq i,j}z_k^{l_k}\Bigg)(z_i+z_j)(z_iz_j)^{\min(l_i,l_j)}
\sign(l_i-l_j)\,\frac{z_i^{|l_i-l_j|}-z_j^{|l_i-l_j|}}{z_i-z_j}\\
&=\Bigg(\prod_{k\neq i,j}z_k^{l_k}\Bigg)(z_i+z_j)(z_iz_j)^{\min(l_i,l_j)}
\sign(l_i-l_j)\sum_{k=0}^{|l_i-l_j|-1}z_i^{|l_i-l_j|-1-k}z_j^k\,,
\end{aligned}\label{inRn}
\end{equation}
where
$$
\sign(p)=
  \begin{cases}
   -1,\; & p<0, \\
    0,\; & p=0, \\
    1,\; & p>0.
  \end{cases}
$$
The resulting polynomial thus belongs to $\cR_n$. Indeed, it is a
linear combination of monomials $\prod_k z_k^{\tl_k}$ with
$\tl_k=l_k$ for $k\neq i,j$, and $\tl_i,\tl_j\leq\max(l_i,l_j)$.
Likewise, the second inclusion in \eqref{inclusions} follows from the
identity
\begin{equation}
\begin{aligned}
\frac{z_i-z_j}{z_i+z_j}\,(1-\tK_{ij})\prod_k z_k^{l_k}
&=\Bigg(\prod_{k\neq
i,j}z_k^{l_k}\Bigg)(z_i-z_j)(z_iz_j)^{\min(l_i,l_j)}
s(l_i,l_j)\,\frac{z_j^{|l_i-l_j|}-(-1)^{l_i+l_j}z_i^{|l_i-l_j|}}{z_i+z_j}\\
&=\Bigg(\prod_{k\neq
i,j}z_k^{l_k}\Bigg)(z_i-z_j)(z_iz_j)^{\min(l_i,l_j)}
s(l_i,l_j)\sum_{k=0}^{|l_i-l_j|-1}(-z_i)^{|l_i-l_j|-1-k}z_j^k\,,
\end{aligned}\label{inRn2}
\end{equation}
where
$$
s(p,q)=
  \begin{cases}
    1,\; & p<q,\\
    0,\; & p=q, \\
    -(-1)^{p+q},\; & p>q.
  \end{cases}
$$

We omit the analogous proof for the operators $J_i^-$. Unlike the
previous types of Dunkl operators, the operators
$J_i^+$ in Eq.~\eqref{J+} do not preserve
the polynomial spaces $\cP_n$ and $\cR_k$ with $k\neq m$.
However, the space $\cR_m$ is invariant under the action of
$J_i^+$. In fact, the operator
$$
z_i^2\frac\pa{\pa z_i}-m\,z_i-b'z_i\big(1-(-1)^m K_i\big),
$$
preserves $\cR_m$, and, just as we did in the case of $J_i^0$, one
can show that both
$$
\frac{z_iz_j}{z_i-z_j}\,(1-K_{ij})
\qquad\text{and}\qquad
\frac{z_iz_j}{z_i+z_j}\,(1-\tK_{ij})
$$
preserve $\cR_n$ for any nonnegative integer $n$.

\section{$BC_N$-type spin many-body Hamiltonians}\label{sec.ham}

In Section~\ref{sec.dunkl} we have shown that all three
sets of $B_N$-type Dunkl operators~\eqref{J-}--\eqref{J+}
preserve the finite-dimensional polynomial space $\cR_m$.
In this section we shall use this fundamental property
to construct several families of (Q)ES
many-body Hamiltonians with internal degrees of freedom.

Let $\fS=\myspan\big\{\,|s_1,\dots,s_N\rangle\mid s_i=-M,-M+1,\dots,M;\;
M\in\frac12\NN\,\big\}$ be the Hilbert space of the particles'
internal degrees of freedom or ``spin''. We shall denote by
$S_{ij}$ and $S_i$, $i,j=1,\dots,N$, the spin
permutation and spin reversing operators, respectively, whose
action on a spin state $|s_1,\dots,s_N\rangle$ is defined by
\begin{equation}
\begin{aligned}
& S_{ij}|s_1,\dots,s_i,\dots,s_j,\dots,s_N\rangle=|s_1,\dots,
s_j,\dots,s_i,\dots,s_N\rangle\,,\\
& S_i|s_1,\dots,s_i,\dots,s_N\rangle=|s_1,\dots,-s_i,\dots,s_N\rangle\,.
\label{SS}
\end{aligned}
\end{equation}
The operators $S_{ij}$ and $S_i$ are represented in
$\fS$ by $(2M+1)^N$-dimensional Hermitian
matrices, and obey identities analogous to
\eqref{KRalg}. The notation $\tS_{ij}=S_iS_jS_{ij}$ shall also be used
in what follows.

We shall deal in this paper with a system of
$N$ identical fermions, so that the physical states
are completely antisymmetric under permutations of the particles.
A physical state $\psi$ must therefore satisfy $\La_0\psi=\psi$, where
$\La_0$ is the antisymmetrisation operator defined by the relations
$\La_0^2=\La_0$ and $\Pi_{ij}\La_0=-\La_0$, $j>i=1,\ldots,N$,
with $\Pi_{ij}=K_{ij}S_{ij}$. Since $K_{ij}^2=1$,
the above relations are equivalent to $K_{ij}\La_0=-S_{ij}\La_0$,
$j>i=1,\ldots,N$. For the lowest values of $N$, the antisymmetriser $\La_0$ is
given by
\begin{align*}
& N=2:\qquad \La_0=\frac12\,\big(1-\Pi_{12}\big)\,,\\[1mm]
& N=3:\qquad
\La_0=\frac16\,\big(1-\Pi_{12}-\Pi_{13}-\Pi_{23}+\Pi_{12}\Pi_{13}+
\Pi_{12}\Pi_{23}\big)\,.
\end{align*}

Our aim is to construct new (Q)ES Hamiltonians symmetric under the
Weyl group of type $B_N$ generated by the permutation
operators $\Pi_{ij}$ and the sign reversing operators $K_iS_i$.
The corresponding algebraic eigenfunctions will
be antisymmetric under a change of sign
of both the spatial and spin variables of any particle, and therefore satisfy
$\La\psi=\psi$, where $\La$ is the projection on states antisymmetric
under permutations and sign reversals. The total antisymmetriser $\La$ is
determined by the relations $\La^2=\La$ and
\begin{equation}
    K_{ij}\La=-S_{ij}\La,\qquad K_i\La=-S_i\La,\qquad
    j>i=1,\ldots,N.
    \label{La}
\end{equation}
It may be easily shown that
$$
\La=\frac1{2^N}\bigg(\prod_i (1-K_iS_i)\bigg)\La_0\,.
$$

Following closely the procedure outlined in~\cite{FGGRZ01},
we shall consider a quadratic combination of the Dunkl
operators~\eqref{J-}--\eqref{J+} of the form
\begin{equation}\label{H*}
  -H^\ast=\sum_i \Big(c_{++}(J_i^+)^2+c_{00}(J_i^0)^2+c_{--}(J_i^-)^2
  +c_0\,J_i^0\Big)\,,
\end{equation}
where $c_{++},c_{00},c_{--},c_0$ are arbitrary real constants such
that $c_{++}^2+c_{00}^2+c_{--}^2\neq 0$. The second-order differential-difference
operator~\eqref{H*} possesses the following remarkable properties.
First, it is a quasi-exactly solvable operator, since it leaves
invariant the polynomial space $\cR_m$. In particular, if $c_{++}=0$
the operator $H^\ast$ preserves $\cR_n$ (and $\cP_n$) for {\em any} nonnegative
integer $n$, and is therefore exactly solvable. Secondly,
$H^\ast$ commutes with $K_{ij}$, $K_i$, $S_{ij}$, and $S_i$ for all
$i,j=1,\dots,N$. This follows immediately from the commutation
relations~\eqref{commJKR}. Note that none of the terms
$$
\sum_i [J_i^\pm,J_i^0]\,,\qquad \sum_i \{J_i^\pm,J_i^0\}\,,\qquad
\sum_i J_i^\pm
$$
commute with $K_i$, and for that reason they have not been included in
the definition of $H^\ast$. We have also discarded the term
$\sum_i\{J_i^+,J_i^-\}$ because it differs from
$2\sum_i \big[(J_i^0)^2+(b'-b)J_i^0\big]$ by a constant operator.

Since $H^\ast$ preserves the polynomial module $\cR_m$, commutes
with $\La$, and acts trivially on $\fS$, the module
\begin{equation}\label{spinmod}
  \BcR_m=\La\big(\cR_m\otimes\fS\big)
\end{equation}
is also invariant under $H^\ast$. It follows from Eqs.~\eqref{La}
that the action of the operators $K_{ij}$ and $K_i$ on the
module $\BcR_m$ coincides with that of the spin operators
$-S_{ij}$ and $-S_i$, respectively. Therefore, the differential
operator $\BH$ obtained from $H^\ast$ by the formal substitutions
$K_{ij}\ra-S_{ij}$, $K_i\ra-S_i$, $i,j=1,\dots,N$, also preserves
the module $\BcR_m$. For the same reason, if the coefficient $c_{++}$
in Eq.~\eqref{H*} vanishes, the operator $\BH$ leaves the
modules $\BcR_n$ and $\BcP_n=\La\big(\cP_n\otimes\fS\big)$ invariant for any non-negative
integer $n$. Using the formulae~\eqref{J-}--\eqref{J+} in the
Appendix for the squares of the Dunkl operators, we get the following
explicit expression for the {\em gauge spin Hamiltonian} $\BH$:
\begin{align}\label{BH}
-\BH
=&\sum_i\Big(P(z_i)\pa^2_{z_i}+Q(z_i)\pa_{z_i}+R(z_i)\Big)
+4a\sum_{i\neq j}\,\frac{z_iP(z_i)}{z_i^2-z_j^2}\,\pa_{z_i}
-\sum_i\bigg(\frac{b\,c_{--}}{z_i^2}\,(1+S_i)
+b'c_{++}z_i^2\big(1+(-1)^mS_i\big)\bigg)\notag\\
{}-a&\sum_{i\neq j}P(z_i)\bigg(\frac{1+S_{ij}}{(z_i-z_j)^2}
+\frac{1+\tS_{ij}}{(z_i+z_j)^2}\bigg)
+\frac{a\,c_{++}}2\,\sum_{i\neq j}\Big((z_i+z_j)^2(1+S_{ij})
+(z_i-z_j)^2(1+\tS_{ij})\Big)+\BC\,,
\end{align}
where
\begin{equation}\label{PQRC}
\begin{aligned}
&P(z)=c_{++}z^4+c_{00}z^2+c_{--}\,,\\
&Q(z)=2c_{++}\big(1-m-b'+2a(1-N)\big)z^3
+\Big(c_0+c_{00}\big(1-m+2a(1-N)\big)\Big)z+\frac{2bc_{--}}z\,,\\
&R(z)=c_{++}m(m-1+2b')z^2\,,\\
&\BC=c_{00}\Bigg[\frac{Nm^2}4+\frac{a^2}{12}
\Bigg({\sum_{i,j,k}}'\big[4-(S_{ij}+\tS_{ij})(S_{ik}+\tS_{ik})\big]
+6\sum_{i\neq j}(1-S_iS_j)\Bigg)
+\frac a2\,\sum_{i\neq j}(2+S_{ij}+\tS_{ij})\Bigg]-\frac{Nmc_0}2\,.
\end{aligned}
\end{equation}
Hereafter, the symbol $\sum_{i,j,k}'$ denotes summation in $i,j,k$ with
$i\neq j\neq k\neq i$.

One of the main ingredients of our method is the fact that the
gauge spin Hamiltonian $\BH$ can be reduced to
a {\em physical} spin Hamiltonian
\begin{equation}\label{H}
H=-\sum_i \pa_{x_i}^2+V(\bx)\,,
\end{equation}
where $V(\bx)$ is a Hermitian matrix-valued function, by a suitable
change of variables $\bz=\bze(\bx)$, $\bx=(x_1,\dots,x_N)$ and a
gauge transformation with a scalar function $\mu(\bx)$, namely
\begin{equation}\label{gauget}
\mu\cdot\BH\big|_{\bz=\bze(\bx)}\cdot\mu^{-1}=H\,.
\end{equation}
We emphasize that in general there is no (matrix or
scalar) gauge factor and change of coordinates reducing a given matrix
second-order differential operator in $N$ variables to a physical
Hamiltonian of the form~\eqref{H}; see~\cite{GKO94,FK97} and
references therein for more details. The quadratic combination
$H^\ast$ has precisely been chosen so that such a gauge factor and
change of variables can be easily found for $\BH$. For instance,
we have omitted the otherwise valid term
$\sum_i[J_i^+,J_i^-]$ because it involves first-order derivatives with
matrix-valued coefficients, which are usually very difficult to gauge
away. The gauge factor $\mu$ and change of variables $\bz=\bze(\bx)$
in Eq.~\eqref{gauget} are respectively given by
\begin{equation}\label{gt}
\mu=\exp\bigg(\sum_i\int^{z_i}\frac{Q(y_i)}{2P(y_i)}\,\diff y_i\bigg)
\prod_{i<j}(z_i^2-z_j^2)^a\,\prod_i\,P(z_i)^{-\frac14},
\end{equation}
and
\begin{equation}\label{cov}
  x_i=\ze^{-1}(z_i)=\int^{z_i}\!\frac{\diff y}{\sqrt{P(y)}}\,,\qquad i=1,\dots,N.
\end{equation}
The physical spin potential $V$ reads
\begin{equation}\label{V}
\begin{aligned}
V&=a\sum_{i\neq j}\Bigg[P(z_i)
\bigg(\frac{a+S_{ij}}{(z_i-z_j)^2}+\frac{a+\tS_{ij}}{(z_i+z_j)^2}\bigg)
-\frac{c_{++}}2\Big((z_i+z_j)^2(1+S_{ij})+(z_i-z_j)^2(1+\tS_{ij})\Big)\Bigg]\\
&\quad+\sum_i\bigg[b'c_{++}\,z_i^2\big(1+(-1)^m S_i\big)
+\frac{b\,c_{--}}{z_i^2}\,(1+S_i)+W(z_i)\bigg]+C\,,
\end{aligned}
\end{equation}
where
\begin{equation}\label{U}
W(z)=\frac12\bigg(Q'-\frac{P''}2\bigg)
+\frac1{4P}\bigg(Q-\frac{P'}2\bigg)\bigg(Q-\frac{3P'}2\bigg)+c\,z^2\,,
\end{equation}
and
\begin{equation}\label{cC}
\begin{aligned}
& C=aN(N-1)\Big[c_0-\frac{c_{00}}3\,\big(a(2N-1)+3(m-1)\big)\Big]-\BC\,,\\
& c=-c_{++}\Big(2a^2(N-1)(2N-1) +4a(N-1)(b'+m-1)+m(2b'+m-1)\Big)\,.
\end{aligned}
\end{equation}
Note that the change of variables~\eqref{cov},
and hence the potential $V(\bx)$, are defined up to an arbitrary
translation in {\em each} coordinate $x_i$, $i=1,\dots,N$.
The hermiticity of the potential~\eqref{V} is a consequence of the Hermitian
character of the spin operators $S_{ij}$ and $S_i$.

The invariance of the module $\BcR_m$ under the gauge spin Hamiltonian
$\BH$ and Eq.~\eqref{gauget} imply that the finite-dimensional
module
\begin{equation}\label{Mm}
    \cM_m=\mu(\bx)\,\La\Big(\cR_m\big(\bze(\bx)\big)\otimes\fS\Big).
\end{equation}
is invariant under the physical spin Hamiltonian $H$.
Therefore, any quadratic combination $H^\ast$ of the form~\eqref{H*}
leads to a (quasi-)exactly solvable spin many-body
potential~\eqref{V}--\eqref{cC} (provided of course that the module $\cM_m$
is not trivial). In particular, if the coefficient $c_{++}$ vanishes,
the spin Hamiltonian $H$ with potential~\eqref{V} is exactly solvable,
since it leaves invariant the infinite chains of finite-dimensional
modules $\cM_n$ and $\cN_n=\mu(\bx)\,
\La\Big(\cP_n\big(\bze(\bx)\big)\otimes\fS\Big)$, $n\in\NN$\,.

Our goal is to obtain a complete classification of the
(Q)ES spin potentials of the form~\eqref{V}--\eqref{cC}.
The key observation used to perform this classification is
the fact that different gauge spin Hamiltonians $\BH$ may yield the
same physical potential. This follows from the form invariance of the
linear spaces $\myspan\{J_i^-(\bz),J_i^0(\bz),J_i^+(\bz)\}$, $i=1,\dots,N$,
under projective (gauge) and scale transformations, given respectively by
\begin{alignat}{2}\label{proj}
& z_j\:\mapsto\: w_j=\frac1{z_j}\,,&\qquad
&J_i^\ep(\bz)\:\mapsto\:\widetilde{J_i^\ep}(\bw)
=\Big(\prod_j z_j^{-m}\Big)J_i^\ep(\bz)\Big(\prod_j z_j^m\Big),
\qquad j=1,\dots,N,\quad \ep=\pm,0,
\\
\intertext{and}
\label{scale}
& z_j\:\mapsto\: w_j={\la z_j}\,,&\qquad
&J_i^\ep(\bz)\:\mapsto\:\widetilde{J_i^\ep}(\bw)=J_i^\ep(\bz),\qquad
j=1,\dots,N,\quad\ep=\pm,0,
\end{alignat}
where $\la\ne0$ is real or purely imaginary. Indeed, we get
$$
\widetilde{J_i^-}(\bw)=\left.-J_i^+(\bw)\right|_{b'\to b}\,,
\qquad
\widetilde{J_i^0}(\bw)=-J_i^0(\bw)\,,
\qquad
\widetilde{J_i^+}(\bw)=\left.-J_i^-(\bw)\right|_{b\to b'}\,,
$$
for the projective transformations~\eqref{proj}, and
$\widetilde{J_i^\ep}(\bw)=\la^{-\ep}J_i^\ep(\bw)$ for the scale
transformations~\eqref{scale}. This implies that the resulting
quadratic combination $\tH^\ast$ is still of the form~\eqref{H*}, with
(in general) different coefficients $\tc_{++}$, $\tc_{00}$,
$\tc_{--}$, and $\tc_0$. Using these transformations, we can reduce
the polynomial $P(z)$ in~\eqref{BH} to one of the following seven
canonical forms:
\begin{equation}\label{canon}
\begin{aligned}
& 1.\quad 1\,,\qquad\qquad &
   & 5.\quad \nu(e^{2\iu\theta}-z^2)(e^{-2\iu\theta}-z^2)\,,\\
& 2.\quad \pm\nu z^2\,,\qquad\qquad &
   & 6.\quad \pm\nu(1-z^2)(1-k^2z^2)\,,\\
& 3.\quad \pm\nu (1+z^2)\,,\qquad\qquad &
   & 7.\quad \nu(1-z^2)(1-k^2+k^2z^2)\,,\\
& 4.\quad \pm\nu(1-z^2)^2\,,\qquad\qquad & &
\end{aligned}
\end{equation}
where $\nu>0$, $0<k<1$, and $0<\theta\le\pi/4$.
\section{Classification of QES spin Calogero--Sutherland models}\label{sec.class}
We present in this section the complete classification of all the
(Q)ES spin Calogero--Sutherland models that can be constructed
applying the procedure described in the previous sections. To further
simplify the classification, we note that the scaling
$(c_{\ep\ep},c_0)\mapsto(\la c_{\ep\ep},\la c_0)$ induces the
mapping
\begin{equation}
    V(\bx\,;c_{\ep\ep},c_0)\;\mapsto\;
    V(\bx\,;\la\,c_{\ep\ep},\la\,c_0)
    =\la\,V(\sqrt\la\,\bx\,;c_{\ep\ep},c_0)
    \label{potsc}
\end{equation}
of the corresponding potentials. For this reason, in Cases 2--7 we
shall only list the potential for a suitably chosen value of the parameter
$\nu$. Note furthermore that in Cases 2--4 and 6, once the potential
has been computed for a positive value $\nu_0$ of the parameter $\nu$,
its counterpart for the opposite value $\nu=-\nu_0$ can be immediately
obtained using \eqref{potsc}, namely
$$
V(\bx\,;-\nu_0,c_0) =-V(\iu\kern1pt\bx\,;\nu_0,-c_0)\,.
$$

For the models constructed to be symmetric under the Weyl group of
type $B_N$ spanned by the operators $K_{ij}S_{ij}$ and $K_i S_i$,
$1\le i<j\le N$, the change of variables $\bz = \bze(\bx)$ should be
an odd function of $\bx$, since only in this case $\bx\mapsto-\bx$
corresponds to $\bz\mapsto-\bz$. In all cases except the second one,
this has essentially the effect of fixing the arbitrary constants on which the
change of variables \eqref{cov} depends. For example, in Case 7 with
$\nu=4$ the change of variables is of the form $z_i =
\pm\cn(2x_i+\xi_i\mid k)\equiv \pm\cn(2x_i+\xi_i)$, $1\le i\le N$.
Imposing that $z_i$ be an odd function of $x_i$ for all $i$ and using
the identity
$$
\cn(2x_i+\xi_i)+\cn(-2x_i+\xi_i)=\frac{2\cn \xi_i\,\cn
(2x_i)}{1-k^2\sn^2\xi_i\,\sn^2(2x_i)}\,,
$$
we obtain the condition
$$
\xi_i = (2l_i-1)K\,,\qquad l_i\in\ZZ\,,\quad 1\le i\le N\,,
$$
where $K\equiv K(k)$ is the complete elliptic integral of the first kind
$$
K(k)=\int_0^{\frac \pi2}\frac{\diff\te}{\sqrt{1-k^2\sin^2\te}}\,.
$$
Since $\cn(2x_i-K+2l_i K)=(-1)^{l_i}\cn(2x_i-K)$, symmetry under
exchange of the particles requires that $l_i$ be independent of $i$,
so that $z_i = \pm\cn(2x_i-K)$ for all $i=1,\dots,N$. Taking into
account that both the potential $V$ and the gauge function $\mu$ are
even functions of $\bz$ by Eqs.~\eqref{gt}--\eqref{U}, we
see that the change of variables in this case can be taken as $z_i =
\cn(2x_i-K)$, $1\le i\le N$.

In the classification that follows, we have routinely discarded
constant operators of the form
\begin{equation}\label{V0}
V_0=\ga_0+\ga_1\sum_i S_i +
\ga_2\sum_{i<j} S_i S_j +
\ga_3\sum_{i<j}(S_{ij}+\tS_{ij}) +
\ga_4{\sum_{i,j,k}}'(S_{ij}+\tS_{ij})(S_{ik}+\tS_{ik})\,,\qquad
\ga_i\in\mathbb{R}.
\end{equation}
This is justified, since the operator $V_0$ commutes with $\La$
(it actually commutes with $K_{ij}$, $S_{ij}$, $K_i$, and $S_i$ for
$1\leq i<j\leq N$) and therefore preserves the spaces $\cM_n$
and $\cN_n$ for all $n$.

All the potentials in the classification presented below are singular
on the hyperplanes $x_i=x_j$, $1\le i<j\le N$, where they diverge as
$(x_i-x_j)^{-2}$. In some cases there may be other singular hyperplanes,
near which the potential behaves as the inverse squared distance
to the hyperplane. We shall accordingly choose as domain of
the functions in the Hilbert space of the system a maximal open subset
$X$ of the open set
\begin{equation}
    x_N<x_{N-1}<\dots<x_2<x_1
    \label{confsp}
\end{equation}
containing no singularities of the potential. In all cases except Case
2b, we shall take as boundary conditions defining the eigenfunctions
of $H$ their square integrability on the region $X$ and their
vanishing on the boundary $\pa X$ of $X$ faster than the square root
of the distance to the boundary. Since the algebraic eigenfunctions
that we shall construct are in all cases regular inside $X$, when this
set is bounded the square integrability of the algebraic
eigenfunctions on $X$ is an automatic consequence of their vanishing
on $\pa X$. In Case 2b, the potential is regular and periodic in each
coordinate in an unbounded domain. Therefore the square integrability
of the eigenfunctions should be replaced by a Bloch-type boundary
condition in this case.

For each of the potentials in the classification, we shall list the
domain chosen for its eigenfunctions and the restrictions imposed by
the boundary conditions discussed above on the parameters on which the
potential depends. In particular, the singularity of the potential at
$x_i=x_j$, $1\le i<j\le N$, forces the parameter $a$ to be greater
than $1/2$. Similarly, in all cases except for the second one the
potential is also singular on the hyperplanes $x_i=0$, $1\le i\le N$,
and the vanishing of the algebraic eigenfunctions on these hyperplanes
as $\abs{x_i}^{\frac12+\de}$ with $\de>0$ requires that $b>1/2$. The
conditions
$$
a>\frac12\,,\qquad b>\frac12
$$
shall therefore be understood to hold in all cases. For similar reasons,
in Cases 4b, 5, and 6b we must also have
$$
b'>\frac12\,.
$$
The
potential in each case will be expressed as
$$
V(\bx) = V_{\text{spin}}(\bx)+\sum_i U(x_i)\,,
$$
where the last term, which does not contain the spin operators
$S_{ij}$ and $S_i$, can be viewed as the contribution
of a scalar external field.

We shall use in the rest of this section the
convenient abbreviations
$$
x_{ij}^\pm=x_i\pm x_j\,,\qquad
\al=a(N-1)+\frac 12\,(b+b'+m)\,.
$$
\\
\smallskip
\nid{\bf Case 1.}\quad $\displaystyle P(z)=1$.\\[3pt]
{\em Change of variables}:\quad $\displaystyle{z=x}$.\\[3pt]
{\em Gauge factor}:
\begin{equation}\label{mu1}
  \mu(\bx)=\prod_{i<j}(x_{ij}^-\,x_{ij}^+)^a
  \prod_i x_i^b\,\e^{-\frac12 \om x_i^2}.
\end{equation}
{\em Scalar external potential}:
\begin{equation}\label{U1}
U(x)=\om^2\,x^2\,.
\end{equation}
{\em Spin potential}:
\begin{equation}\label{V1}
V_{\text{spin}}(\bx)=2a\,\sum_{i<j}\left[(x_{ij}^-)^{-2}\,(a+S_{ij})
+(x_{ij}^+)^{-2}\,(a+\tS_{ij})\right]
+b\,\sum_i x_i^{-2}\,(b+S_i)\,.
\end{equation}
{\em Parameters}:\quad
$
\om = -\frac12 c_0>0\,.
$\\[3pt]
{\em Domain}:\enspace $0<x_N<\dots<x_1$.\\

\smallskip
\nid{\bf Case 2a.}\quad $\displaystyle P(z)=4\,z^2$.\\[3pt]
{\em Change of variables}:\quad $\displaystyle
z=e^{2 x}$.\\[3pt]
The most general change of variables in this case is $z_i=\la e^{\pm 2
x_i}$, $1\le i\le N$, but the following formulas are independent of
the choice of sign in the exponent and the value of the constant $\la$.\\[3pt]
{\em Gauge factor}:
\begin{equation}\label{mu2a}
\mu(\bx)=\prod_{i<j}\left[\sinh (2x_{ij}^-)\right]^a\,.
\end{equation}
{\em Scalar external potential}:\quad
$
U(x)=0\,.
$\\[3pt]
{\em Spin potential}:
\begin{equation}\label{V2a}
V_{\text{spin}}(\bx)=2a\,\sum_{i<j}\left[\sinh^{-2} x_{ij}^-\,(a+S_{ij})
-\cosh^{-2} x_{ij}^-\,(a+\tS_{ij})\right]\,.
\end{equation}
{\em Parameters}:\quad
$
c_0=4m\,.
$
\\[3pt]
{\em Domain}:\enspace $x_N<\dots<x_1$.\\

\smallskip
\nid{\bf Case 2b.}\quad $\displaystyle P(z)=-4\,z^2$.\\[3pt]
{\em Change of variables}:\quad $\displaystyle
z=\e^{2\iu x}$.\\[3pt]
Again, the most general change of variables is $z_i=\la e^{\pm 2\iu
x_i}$, $1\le i\le N$, but the following formulas do not change when
this is taken into account.\\[3pt]
{\em Gauge factor}:
\begin{equation}\label{mu2b}
\mu(\bx)=\prod_{i<j}\left[\sin (2x_{ij}^-)\right]^a\,.
\end{equation}
{\em Scalar external potential}:\quad
$
U(x)=0\,.
$\\[3pt]
{\em Spin potential}:
\begin{equation}\label{V2b}
V_{\text{spin}}(\bx)=2a\,\sum_{i<j}\left[\sin^{-2} x_{ij}^-\,(a+S_{ij})
+\cos^{-2} x_{ij}^-\,(a+\tS_{ij})\right].
\end{equation}
{\em Parameters}:\quad $ c_0=-4m\,. $\\[3pt] {\em Domain}:\enspace
$x_N<\dots<x_1<x_N+\frac\pi2$.\\[3pt]
\nid Both potentials in this case are invariant under a simultaneous
translation of all the particles' coordinates. The choice $c_0 = \pm
4m$, which simplifies the form of the gauge factor, amounts to fixing
the center of mass energy of the system. Note also that the potentials
in this case do \emph{not} possess $B_N$ symmetry, due to the fact
that the change of variables cannot be made an odd function of $x$ for
any choice of the arbitrary constants. In fact, the sign change
$z_k\mapsto -z_k$ corresponds to the translation $x_k\mapsto
x_k+\iu\kern 1pt\pi/2$ or $x_k\mapsto x_k+\pi/2$, which (as any
overall translation) leaves the potential invariant. The potentials in
this case are therefore best interpreted as $A_N$-type potentials
depending both on spin permutation and sign reversing operators.

For the hyperbolic potential 2a, none of the algebraic formal
eigenfunctions are true eigenfunctions, since they are not square
integrable on their domain. On the other hand, the algebraic
eigenfunctions of the periodic potential 2b are clearly periodic in
each coordinate and regular on their domain, and thus qualify as true
eigenfunctions.\\

\smallskip
\nid{\bf Case 3a.}\quad $\displaystyle P(z)=4\,(1+z^2)$.\\[3pt]
{\em Change of variables}:\quad $\displaystyle z=\sinh (2x)$.\\[3pt]
{\em Gauge factor}:
\begin{equation}\label{mu3a}
\mu(\bx)=\prod_{i<j}\Big[\sinh (2x_{ij}^-)\sinh (2x_{ij}^+)\Big]^a\,
\prod_i \big[\sinh(2x_i)\big]^b \big[\cosh(2x_i)\big]^\be\,.
\end{equation}
{\em Scalar external potential}:
\begin{equation}\label{U3a}
U(x)=-4\be(\be-1)\cosh^{-2} (2x)\,.
\end{equation}
{\em Spin potential}:
\begin{equation}\label{V3a}
\begin{aligned}
V_{\text{spin}}(\bx)=2a\,&\sum_{i<j}\left[\big(\sinh^{-2} x_{ij}^-
-\cosh^{-2} x_{ij}^+\big)\,(a+S_{ij})+
\big(\sinh^{-2} x_{ij}^+
-\cosh^{-2} x_{ij}^-\big)\,(a+\tS_{ij})\right]\\
& {}+4b\,\sum_i \sinh^{-2}(2x_i)\,(b+S_i)\,.
\end{aligned}
\end{equation}
{\em Parameters}:\quad
$\displaystyle
\be=\frac{c_0}8-\Big(a(N-1)+b+\frac m2\Big)<-\big(2a(N-1)+b+m\big)\,.
$
\\[3pt]
{\em Domain}:\enspace $0<x_N<\dots<x_1$.\\

\smallskip
\nid{\bf Case 3b.}\quad $\displaystyle P(z)=-4\,(1+z^2)$.\\[3pt]
{\em Change of variables}:\quad $\displaystyle z=\iu\kern 1pt\sin (2x)$.\\[3pt]
{\em Gauge factor}:
\begin{equation}\label{mu3b}
\mu(\bx)=\prod_{i<j}\Big[\sin (2x_{ij}^-)\sin (2x_{ij}^+)\Big]^a\,
\prod_i \big[\sin(2x_i)\big]^b \big[\cos(2x_i)\big]^\be\,.
\end{equation}
{\em Scalar external potential}:
\begin{equation}\label{U3b}
U(x)=4\be(\be-1)\cos^{-2} (2x)\,.
\end{equation}
{\em Spin potential}:
\begin{equation}\label{V3b}
V_{\text{spin}}(\bx)=2a\,\sum_{i<j}\left[\big(\sin^{-2} x_{ij}^-
+\cos^{-2} x_{ij}^+\big)\,(a+S_{ij})+
\big(\sin^{-2} x_{ij}^+
+\cos^{-2} x_{ij}^-\big)\,(a+\tS_{ij})\right]
+4b\,\sum_i \sin^{-2}(2x_i)\,(b+S_i)\,.
\end{equation}
{\em Parameters}:\quad
$\displaystyle
\be=-\left(\frac{c_0}8+a(N-1)+b+\frac m2\right)>\frac12
$
or $\be=0$.
\\[3pt]
{\em Domain}:\enspace $0<x_N<\dots<x_1<\frac\pi4$,\enspace if
$\be>\frac12$ and
$\be\ne1$;\quad
$0<x_N<\dots<x_1<\frac\pi2-x_2$,\enspace if $\be=0,1$.\\

\smallskip
\nid{\bf Case 4a.}\quad $\displaystyle P(z)=(1-z^2)^2$.\\[3pt]
{\em Change of variables}:\quad $\displaystyle z=\tanh x$.\\[3pt]
{\em Gauge factor}:
\begin{equation}\label{mu4a}
\mu(\bx)=\prod_{i<j}\Big(\sinh x_{ij}^-\sinh x_{ij}^+\Big)^a\,
\prod_i \e^{\be\cosh(2x_i)}\,(\sinh x_i)^b (\cosh x_i)^{b'+m}\,.
\end{equation}
{\em Scalar external potential}:
\begin{equation}\label{U4a}
U(x)=2\be^2 \cosh(4x)+4\be(1+2\al)\,\cosh(2x)\,.
\end{equation}
{\em Spin potential}:
\begin{multline}\label{V4a}
V_{\text{spin}}(\bx)=2a\,\sum_{i<j}\left[\sinh^{-2} x_{ij}^-\,(a+S_{ij})+
\sinh^{-2} x_{ij}^+\,(a+\tS_{ij})\right]\\
+b\,\sum_i \sinh^{-2}\!x_i\,(b+S_i)
-b'\,\sum_i \cosh^{-2}\!x_i\,\big(b'+(-1)^m S_i\big)\,.
\end{multline}
{\em Parameters}:\quad
$
\be=\displaystyle\frac 18\,\big(c_0+2(b-b')\big)<0\,,
$
\enspace or \enspace $\be=0$\, and \,$\al<0$.\\[3pt]
{\em Domain}:\enspace $0<x_N<\dots<x_1$.\\[3pt]
Alternatively, we could have taken the change of variables as
$z=\tanh\big(x-\frac{\iu\pi}2\big)=\coth x$. The gauge factor,
external potential and spin potential become, respectively,
\begin{align}
\mu(\bx)&=\prod_{i<j}\Big(\sinh x_{ij}^-\sinh x_{ij}^+\Big)^a\,
\prod_i \e^{-\be\cosh(2x_i)}\,(\cosh x_i)^b (\sinh
x_i)^{b'+m}\,,\label{mu4abis}\\
U(x)&=2\be^2 \cosh(4x)-4\be(1+2\al)\,\cosh(2x)\,,\label{U4abis}
\end{align}
and
\begin{multline}
V_{\text{spin}}(\bx)=2a\,\sum_{i<j}\left[\sinh^{-2} x_{ij}^-\,(a+S_{ij})+
\sinh^{-2} x_{ij}^+\,(a+\tS_{ij})\right]\\
-b\,\sum_i \cosh^{-2}\!x_i\,(b+S_i)
+b'\,\sum_i \sinh^{-2}\!x_i\,\big(b'+(-1)^m S_i\big)\,.\label{V4abis}
\end{multline}
\\

\nid{\bf Case 4b.}\quad $\displaystyle P(z)=-(1-z^2)^2$.\\[3pt]
{\em Change of variables}:\quad $\displaystyle z=\iu\kern 1pt\tan x$.\\[3pt]
{\em Gauge factor}:
\begin{equation}\label{mu4b}
\mu(\bx)=\prod_{i<j}\Big(\sin x_{ij}^-\sin x_{ij}^+\Big)^a\,
\prod_i \e^{\be\cos(2x_i)}\,(\sin x_i)^b (\cos x_i)^{b'+m}\,.
\end{equation}
{\em Scalar external potential}:
\begin{equation}\label{U4b}
U(x)=-2\be^2 \cos(4x)-4\be(1+2\al)\,\cos(2x)\,.
\end{equation}
{\em Spin potential}:
\begin{multline}\label{V4b}
V_{\text{spin}}(\bx)=2a\,\sum_{i<j}\left[\sin^{-2} x_{ij}^-\,(a+S_{ij})+
\sin^{-2} x_{ij}^+\,(a+\tS_{ij})\right]\\
+b\,\sum_i \sin^{-2}\!x_i\,(b+S_i)
+b'\,\sum_i\cos^{-2}\!x_i\,\big(b'+(-1)^m S_i\big)\,.
\end{multline}
{\em Parameters}:\quad
$
\be=-\displaystyle\frac 18\,\big(c_0+2(b'-b)\big)\,.
$
\\[3pt]
{\em Domain}:\enspace $0<x_N<\dots<x_1<\frac\pi2$.\\[3pt]
The change of variable can also be taken as
$z=\iu\tan\big(x-\frac\pi2\big) =\iu\kern 1pt\cot x$. Since this is the
result of applying an overall \emph{real} translation to the particles'
coordinates, we shall not list the corresponding formulas for the
potential and gauge factor.

\bigskip
\nid{\bf Case 5.}\quad $\displaystyle
P(z)=\big(\e^{2\iu\te}-z^2\big)\big(\e^{-2\iu\te}-z^2\big)$.\\[3pt]
{\em Change of variables}:
$$
z=\frac{\sn x\,\dn
x}{\cn x}\,,
$$
where the modulus of the elliptic functions is $k=\cos\te$. We shall
also use in what follows the customary notation $k'$ for the
complementary modulus $\sqrt{1-k^2}$.\\[3pt]
{\em Gauge factor}:
\begin{multline}\label{mu5}
\mu(\bx)=\prod_{i<j}\left(\frac{\sn x_{ij}^-\dn x_{ij}^- \sn
x_{ij}^+\dn x_{ij}^+}{1-k^2 \sn^2 x_{ij}^-\sn^2
x_{ij}^+}\right)^a
\prod_i
\exp\Big\{\be \arctan\Big[\frac k{k'}\,\cn(2x_i)\Big]\Big\}\\
\times
\big[\sn(2x_i)\big]^b
\big[1+\cn(2x_i)\big]^{\frac 12(b'-b+m)}
\big[\dn(2x_i)\big]^{-\al}\,.
\end{multline}
{\em Scalar external potential}:
\begin{equation}\label{U5}
U(x)=4k'{}^2\dn^{-2}(2x)\left[\be^2-\al(\al+1)-\frac {k
\be}{k'}\,(1+2\al)\,\cn(2x)\right]\,.
\end{equation}
{\em Spin potential}:
\begin{equation}\label{V5}
\begin{aligned}
V_{\text{spin}}(\bx)=2a\,&\sum_{i<j}\left[\left(\frac{\dn^2 x_{ij}^-}{\sn^2
x_{ij}^-}-k^2k'{}^2\,\frac{\sn^2 x_{ij}^+}{\dn^2
x_{ij}^+}\right)(a+S_{ij})+
\left(\frac{\dn^2 x_{ij}^+}{\sn^2
x_{ij}^+} - k^2k'{}^2\,\frac{\sn^2 x_{ij}^-}{\dn^2
x_{ij}^-}\right)(a+\tS_{ij})
\right]\\
{}+b\,&\sum_i \Big(\frac{\cn x_i}{\sn x_i\,\dn x_i}\Big)^2 (b+S_i)
+b'\,\sum_i \Big(\frac{\sn x_i\,\dn x_i}{\cn x_i}\Big)^2 \big(b'+(-1)^m
S_i\big).
\end{aligned}
\end{equation}
{\em Parameters}:\quad
$\displaystyle
\be = -\frac 1{8kk'}\,\Big(c_0+2(k^2-k'^2)(b-b')\Big)\,.
$\\[4pt]
{\em Domain}:\enspace $0<x_N<\dots<x_1<K$.\\

\smallskip\noindent
An alternative form for the change of variables in this case is
$$
z =\frac{\sn(x-K)\,\dn(x-K)}{\cn(x-K)} = -\frac{\cn x}{\sn x\,\dn x}\,.
$$
The resulting potential is obtained from the previous one by applying
the overall real translation $x_i\mapsto x_i-K$, $i=1,\dots,N$. Note
also that, although $z_i$ is singular at the zeros
of $\cn x_i$, the algebraic eigenfunctions satisfy the appropriate
boundary condition on these hyperplanes on account of the inequality $b'>1/2$ and the
identity
$$
1+\cn(2x) = \frac{2\,\cn^2 x}{1-k^2\sn^4 x}\,.
$$

\bigskip
\nid{\bf Case 6a.}\quad $\displaystyle
P(z)=4(1-z^2)(1-k^2\,z^2)$.\\[3pt]
{\em Change of variables}:\quad $\displaystyle{z=\sn (2x)}$.\\[3pt]
Here, as in the remaining cases, the Jacobian elliptic functions have
modulus $k$.\\[3pt]
{\em Gauge factor}:
\begin{equation}\label{mu6a}
\mu(\bx)=\prod_{i<j}\frac{\big(\sn x_{ij}^- \cn x_{ij}^- \dn x_{ij}^- \sn
x_{ij}^+ \cn x_{ij}^+ \dn x_{ij}^+\big)^a}{\big(1-k^2 \sn^2 x_{ij}^-\sn^2
x_{ij}^+\big)^{2a}}\,
\prod_i
\big[\sn(2x_i)\big]^b\,\big[\cn(2x_i)\big]^\be\,\big[\dn(2x_i)\big]^{\be'}\,.
\end{equation}
{\em Scalar external potential}:
\begin{equation}\label{U6a}
U(x)=4k'{}^2\big[\be(\be-1)\,\cn^{-2}(2x)-\be'(\be'-1)\,\dn^{-2}(2x)\big]\,.
\end{equation}
{\em Spin potential}:
\begin{align}\label{V6a}
V_{\text{spin}}(\bx)=2a\,&\sum_{i<j}
\Bigg[
\Bigg(\frac{\cn^2 x_{ij}^- \dn^2
x_{ij}^-}{\sn^2 x_{ij}^-}
+k'{}^4 \frac{\sn^2 x_{ij}^+}{\cn^2 x_{ij}^+ \dn^2
x_{ij}^+}\Bigg) (a+S_{ij})
+\Bigg(\frac{\cn^2 x_{ij}^+ \dn^2
x_{ij}^+}{\sn^2 x_{ij}^+}
+k'{}^4 \frac{\sn^2 x_{ij}^-}{\cn^2 x_{ij}^- \dn^2
x_{ij}^-}\Bigg) (a+\tS_{ij})
\Bigg]\notag\\
{}+4b\,&\sum_i \sn^{-2}(2x_i)\,(b+S_i)
+4k^2b'\,\sum_i \sn^2 (2x_i)\,(b'+(-1)^m S_i)\,.
\end{align}
{\em Parameters}:
$$
\be=-\frac 1{8k'{}^2}\Big(c_0+4(1+k^2)(b-b')\Big)-\al
>\frac12\enspace\text{ or }\enspace\be=0\,,\qquad
\be'=\frac 1{8k'{}^2}\Big(c_0+4(1+k^2)(b-b')\Big)-\al\,.
$$
\\[3pt]
{\em Domain}:\enspace $0<x_N<\dots<x_1<\frac K2$,\enspace if
$\be>\frac12$ and
$\be\ne1$;\quad
$0<x_N<\dots<x_1<K-x_2$,\enspace if $\be=0,1$.\\

\bigskip
\nid{\bf Case 6b.}\quad $\displaystyle P(z)=-4(1-z^2)(1-k'{}^2\,z^2)$.\\[3pt]
{\em Change of variables}:
$$
z=\iu\,\frac{\sn
(2x)}{\cn(2x)}\,.
$$
{\em Gauge factor}:
\begin{equation}\label{mu6b}
\mu(\bx)=\prod_{i<j}\frac{\big(\sn x_{ij}^- \cn x_{ij}^- \dn x_{ij}^- \sn
x_{ij}^+ \cn x_{ij}^+ \dn x_{ij}^+\big)^a}{\big(1-k^2 \sn^2 x_{ij}^-\sn^2
x_{ij}^+\big)^{2a}}\,
\prod_i
\big[\sn(2x_i)\big]^b\,\big[\cn(2x_i)\big]^{b'+m}\,\big[\dn(2x_i)\big]^{\be'}\,.
\end{equation}
{\em Scalar external potential}:
\begin{equation}\label{U6b}
U(x)=4\big[\be(\be-1)\,k^2 \sn^2(2x)-\be'(\be'-1)\,k'{}^2\dn^{-2}(2x)\big]\,.
\end{equation}
{\em Spin potential}:
\begin{align}\label{V6b}
V_{\text{spin}}(\bx)=2a\,&\sum_{i<j}\Bigg[
\Bigg(\frac{\dn^2
x_{ij}^-}{\sn^2 x_{ij}^- \cn^2 x_{ij}^- }
+k^4\,\frac{\sn^2 x_{ij}^+ \cn^2 x_{ij}^+ }{\dn^2
x_{ij}^+}\Bigg) (a+S_{ij})
+ \Bigg(\frac{\dn^2
x_{ij}^+}{\sn^2 x_{ij}^+ \cn^2 x_{ij}^+ }
+k^4\,\frac{\sn^2 x_{ij}^- \cn^2 x_{ij}^- }{\dn^2
x_{ij}^-}\Bigg) (a+\tS_{ij})
\Bigg]\notag\\
{}+4b\,&\sum_i \sn^{-2}(2x_i)\,(b+S_i)+4k'{}^2 b'\,\sum_i \cn^{-2}
(2x_i)\,(b'+(-1)^m S_i)\,.
\end{align}
{\em Parameters}:
$$
\be=\frac 1{8k^2}\Big(c_0+4(1+k'{}^2)(b'-b)\Big)-\al,\qquad
\be'=-\frac 1{8k^2}\Big(c_0+4(1+k'{}^2)(b'-b)\Big)-\al\,.
$$
\\[3pt]
{\em Domain}:\enspace $0<x_N<\dots<x_1<\frac K2$.\\[3pt]
In spite of the singularity of $z_i$ at the zeros of $\cn (2x_i)$, the
vanishing of the gauge factor on these hyperplanes clearly implies
that the algebraic eigenfunctions fulfill the appropriate boundary condition.

\bigskip
\nid{\bf Case 7.}\quad $\displaystyle P(z)=4(1-z^2)(k'{}^2+k^2\,z^2)$.\\
{\em Change of variables}:
$$
z=\cn (2x-K) = k'\,\frac{\sn(2x)}{\dn(2x)}\,.
$$
{\em Gauge factor}:
\begin{equation}\label{mu7}
\mu(\bx)=\prod_{i<j}\frac{\big(\sn x_{ij}^- \cn x_{ij}^- \dn x_{ij}^- \sn
x_{ij}^+ \cn x_{ij}^+ \dn x_{ij}^+\big)^a}{\big(1-k^2 \sn^2 x_{ij}^-\sn^2
x_{ij}^+\big)^{2a}}\,
\prod_i
\big[\sn(2x_i)\big]^b\,\big[\cn(2x_i)\big]^\be\,\big[\dn(2x_i)\big]^{b'+m}\,.
\end{equation}
{\em Scalar external potential}:
\begin{equation}\label{U7}
U(x)=4\big[\be'(\be'-1)\,k^2\sn^2(2x)+\be(\be-1)\,k'{}^2 \cn^{-2}(2x)
\big]\,.
\end{equation}
{\em Spin potential}:
\begin{align}\label{V7}
V_{\text{spin}}(\bx)=2a\,&\sum_{i<j}\Bigg[
\Bigg(\frac{\cn^2 x_{ij}^-
}{\sn^2 x_{ij}^- \dn^2 x_{ij}^-}
+\frac{\sn^2 x_{ij}^+ \dn^2 x_{ij}^+}{\cn^2
x_{ij}^+}\Bigg) (a+S_{ij})
+\Bigg(\frac{\cn^2 x_{ij}^+
}{\sn^2 x_{ij}^+ \dn^2 x_{ij}^+}
+\frac{\sn^2 x_{ij}^- \dn^2 x_{ij}^-}{\cn^2
x_{ij}^-}\Bigg) (a+\tS_{ij})
\Bigg]\notag\\
{}+4b\,&\sum_i\sn^{-2}(2x_i)\,(b+S_i)
-4k'{}^2 b'\,\sum_i\dn^{-2}
(2x_i)\,(b'+(-1)^m S_i)\,. \kern 4em
\end{align}
{\em Parameters}:
$$
\be=-\left(\frac{c_0}8+\frac12(k^2-k'{}^2)(b'-b)+\al\right)>\frac12
\enspace\text{ or }\enspace\be=0\,,\qquad
\be'=\frac{c_0}8+\frac12(k^2-k'{}^2)(b'-b)-\al\,.
$$
\\[3pt]
{\em Domain}:\enspace $0<x_N<\dots<x_1<\frac K2$,\enspace if
$\be>\frac12$ and
$\be\ne1$;\quad
$0<x_N<\dots<x_1<K-x_2$,\enspace if $\be=0,1$.\\

\section{Discussion}\label{sec.disc}

The $BC_N$-type potentials constructed in the previous section can
be expressed in a unified way that we shall now describe. In the first
place, apart from irrelevant constant operators of the form~\eqref{V0},
the spin potential can be written as
\begin{align}
    V_{\text{spin}}(\bx)=
    2a\,&\sum_{i<j}\Big[\big(v(x_{ij}^-)+v(x_{ij}^+ + P_1)\big)(a+S_{ij})
    + \big(v(x_{ij}^+)+v(x_{ij}^- + P_1)\big)(a+\tS_{ij})
    \Big]\notag\\
    {}+b\,&\sum_i\big(v(x_i)+v(x_i+P_1)\big)(b+S_i)
    + b'\,\sum_i\big(v(x_i+P_2)+v(x_i+P_1+P_2)\big)\big(b'+(-1)^m S_i\big),
    \label{Vspin}
\end{align}
where $v$ is a (possibly degenerate) elliptic function, and $P_1$ and
$P_2$ are suitably chosen primitive half-periods of $v$ (see
Table \ref{tab.v}).
\begin{table}[t]
\begin{tabular}{cccc}\toprule
    \Strut \hbox to .25cm{}Case\hbox to .5cm{}& $v(x)$&
    $P_1$& $P_2$\\[1mm]
    \colrule
    1.\Strut &  $x^{-2}$& $\infty$& $\infty$\\[1mm]
    \colrule
    3a.\BStrut & $\sinh^{-2}x$& $\ds\frac{\iu\pi}2$& $\infty$\\[1mm]
    \colrule
    3b.\BStrut & $\sin^{-2}x$& $\ds\frac\pi2$& $\infty$\\[1mm]
    \colrule
    4a.\BStrut & $\sinh^{-2}x$ & $\infty$& $\ds\frac{\iu\pi}2$\\[1mm]
    \colrule
    4b.\BStrut & $\sin^{-2}x$ & $\infty$& $\ds\frac\pi2$\\[1mm]
    \colrule
    5.\BStrut & $\ds\frac{\dn^2 x}{\sn^2 x}$ & \hbox to
    .5cm{}$K+\iu K'$\hbox to .5cm{}& $K$\\[1mm]
    \colrule
    6a.\BStrut & \hbox to .25cm{}$\ds\frac{\cn^2 x\,\dn^2 x}{\sn^2
    x}$\hbox to .25cm{}& $K$& $\ds\frac{\iu K'}2$\\[1mm]
    \colrule
    6b.\BStrut & $\ds\frac{\dn^2 x}{\sn^2 x\,\cn^2 x}$ & $\iu
    K'$& $\ds\frac K2$\\[1mm]
    \colrule
    7.\BStrut & $\ds\frac{\cn^2 x}{\sn^2 x\,\dn^2 x}$ & $K$&
    $\ds\frac12 (K+\iu K')$\\[1mm]
    \botrule
\end{tabular}
\caption{Function $v(x)$ and its primitive half-periods $P_i$
(see Eq.~\eqref{Vspin}) for each of the $BC_N$-type potentials in Section
\ref{sec.class}.}
\label{tab.v}
\end{table}
In particular, when one of the periods $P_i$ of $v$ goes to infinity,
expressions like $v(x+P_i)$ with $x\in\RR$ finite are defined as zero.
In Case 1, both periods are infinite and $v(x+P_1+P_2)$ is also
defined as zero. Furthermore, the constant $K'\equiv K'(k)$ in the
elliptic Cases 5--7 is the complete elliptic integral of the first
kind defined by $K'(k)=K(k')$. Using this notation, it is easy to
verify that in the non-degenerate elliptic Cases 5--7 the scalar
external potential $U(x)$ can be written as
\begin{equation}
    U(x) = \la(\la-1)\Big[v\big(x+\textstyle\frac12
    P_1\big)+v\big(x-\frac12 P_1\big)\Big]+
    \la'(\la'-1)\Big[v\big(x+\frac12 P_1+P_2\big)+ v\big(x-\frac12
    P_1+P_2\big)\Big],
    \label{Ux}
\end{equation}
where $\la=\be$ and $\la'=\be'$ in Cases 6--7, while $\la=-\al+\iu\be$
and $\la'=-\al-\iu\be$ in Case 5.
Formula \eqref{Ux} holds also in Case 3, with $\la=\be$ and $\la'=0$.
In Cases 1 and 4 Eq.~\eqref{Ux} cannot be directly applied, since in
these cases all the terms in \eqref{Ux} are either indeterminate or
zero. However, the potentials in Cases 4a and 4b can be obtained from
that of Case 5 in the limits $\te\to0$ and $\te\to\pi/2$,
respectively. Likewise, applying the rescaling $x_i\mapsto \nu x_i$
($i=1,\dots,N$, $\nu>0$) to the potential of type 3a or 3b one
obtains the potential of type 1 by taking $\be=-\om/(4\nu^2)$ and
letting $\nu\to0$.

The function $v(x)$ that determines the potential $V(\bx)$ in the
elliptic Cases 5--7 according to Eqs.~\eqref{Vspin} and \eqref{Ux} can
be expressed in a systematic way in terms of the Weierstrass function
$\wp(x;\om_1,\om_3)$ with primitive half-periods $\om_1=K$ and
$\om_3=\iu K'$. Indeed, dropping inessential constant operators we
have
\begin{equation}
    v(x) = \ep[\wp(x;\om_1,\om_3) + \wp(x+2P_2;\om_1,\om_3)]\,,
    \label{vp}
\end{equation}
where $P_2$ is the primitive half-period of $v$ listed in Table
\ref{tab.v}, and $\ep=1$ for Cases 6--7 while $\ep=1/2$ for Case 5
(the only case in which $2P_2=2K$ is a period of $\wp$). Since in
Cases 6--7 $P_1$ and $2P_2$ are primitive half-periods of
$\wp$, the well-known second-order modular transformation
of the Weierstrass function \cite{La89} applied to Eq.~\eqref{vp} leads to
the equality
\begin{equation}
    v(x) = \wp(x;P_1,P_2)\,,
    \label{valt}
\end{equation}
where the primitive half-periods $P_1$ and $P_2$ are listed in Table
\ref{tab.v}, and we have dropped an irrelevant additive constant.
Substituting Eq.~\eqref{valt} into Eqs.~\eqref{Vspin} and \eqref{Ux}
and applying once again a modular transformation to the one-particle
terms we readily obtain the following remarkable expression for the
potential $V(\bx)$ in Cases 5--7:
    \begin{align}\label{genpot}
    V(\bx) =
    2a\,&\sum_{i<j}\Big[\big(\wp(x_{ij}^-;P_1,P_2)+\wp(x_{ij}^+ +
    P_1;P_1,P_2)\big)(a+S_{ij}) +
    \big(\wp(x_{ij}^+;P_1,P_2)+\wp(x_{ij}^- +
    P_1;P_1,P_2)\big)(a+\tS_{ij}) \Big]\notag\\
        {}+4b\,&\sum_i\wp(2x_i;P_1,2P_2)(b+S_i)+
        4b'\,\sum_i\wp(2x_i+2P_2;P_1,2P_2)\big(b'+(-1)^m
        S_i\big)\\
        {}+4\,&\sum_i\Big[\la(\la-1)\,\wp(2x_i+P_1;P_1,2P_2)
        +\la'(\la'-1)\,\wp(2x_i+P_1+2P_2;P_1,2P_2)\Big].\notag
    \end{align}
One of the main results in this paper is thus the fact that the
potential \eqref{genpot} is QES provided that the \emph{ordered} pair
$(P_1,P_2)$ is chosen from Cases 5--7 in Table \ref{tab.v}. In fact,
the remaining $BC_N$-type (Q)ES spin potentials listed in
Section~\ref{sec.class} can be obtained from the potentials in
Eqs.~\eqref{genpot} by sending one or both of the half-periods of the
Weierstrass function to infinity. This is of course reminiscent of the
analogous property of the integrable scalar Calogero--Sutherland
models associated to root systems~\cite{OP83}.\\

The potentials in Cases 1, 2, and 3 are ES for all values of the parameters.
(In Case~3, the dependence of the parameter $\be$ on $m$ through $\al$
can be absorbed in the coefficient $c_0$.) The potentials of type 4
are also ES for $\be=0$. The elliptic potentials in Cases 5--7 are always QES.

All the potentials presented in Section~\ref{sec.class} are new, except
for Cases 1~and~4. Case~1 is the rational $B_N$-type model introduced
by Yamamoto~\cite{Ya95} and studied by Dunkl~\cite{Du98}.
Case 4b for $\be=0$ is Yamamoto's $B_N$-type trigonometric potential with $\la_1=-b$
(in the notation of Ref.~\cite{Ya95}), and either $\la_1'=-b'<-1/2$ for $m$ even or
$\la_1'=b'>1/2$ for $m$ odd. Our results thus establish the exact solvability of
the trigonometric Yamamoto model when $\abs{\la_1'}>1/2$.\\

It should be noted that the method developed in this paper
admits a number of straightforward generalizations. In the first
place, the algebraic states could be chosen symmetric under
sign reversals. The resulting Hamiltonians would coincide with
the ones presented in Section~\ref{sec.class} with $S_i$ replaced by $-S_i$.
In particular, if $b=b'=0$ one can obtain algebraic eigenfunctions
of {\em both} types (symmetric and antisymmetric under sign reversals)
for the {\em same} Hamiltonian. The construction can also
be applied to a system of $N$ identical bosons, just by replacing
the antisymmetriser $\La_0$ by the projector on states symmetric
under permutations of the particles. Choosing a system of fermions
is motivated by the fact that the internal degrees of freedom
can be naturally interpreted as the physical spin of the particles
when $M=1/2$.

The procedure described in Section~\ref{sec.ham} relies on the
algebraic identities analogous to~\eqref{KRalg} satisfied by the
spin operators $S_{ij}$ and $S_i$, and not on the particular
realization~\eqref{SS}. For instance, replacing the
operators $S_{ij}$ by new operators $\hat S_{ij}$
spanning one of the anyon-like realizations introduced by
Basu-Mallick~\cite{BM96} would yield further families of
(Q)ES spin Calogero--Sutherland models.

\section{Exact solutions for an elliptic QES model}\label{sec.ex}

As an illustration of the procedure described in the previous sections,
we shall now compute the algebraic sector of the spectrum for
the elliptic QES potential of type 6a in Eqs.~\eqref{U6a}--\eqref{V6a}
in the case of two and three particles of spin $1/2$ ($N=2,3$ and $M=1/2$),
for $m=1,2,3$. Note that in the spin $1/2$ case, the spin
permutation and sign reversing operators $S_{ij}$ and $S_i$
can be expressed in terms of the usual one-particle ${\rm SU}(2)$ spin operators
$\bsi_i=(\si_i^1,\si_i^2,\si_i^3)$ in the more familiar way
\[
S_{ij}=2\,\bsi_i\cdot\bsi_j+\frac12\,,\qquad
S_i=2\si_i^1\,.
\]
The operator $H^\ast$ corresponding to the potential~\eqref{U6a}--\eqref{V6a}
reads
\begin{equation}\label{H*ex}
H^\ast=-\sum_{i=1}^N \Big(4k^2(J_i^+)^2-4(1+k^2)(J_i^0)^2+4(J_i^-)^2
+c_0\,J_i^0\Big)+\BC^\ast+E_0\,,
\end{equation}
where $\BC^\ast$ is the constant operator obtained by
replacing $S_{ij}$ by $-K_{ij}$ and $S_i$ by $-K_i$ in the
expression~\eqref{PQRC} for $\BC$, and the scalar constant $E_0$ is given by
$$
E_0=c_0N\Big(a(N-1)+b'+m+\frac12\Big)
-2N(1+k^2)\Big(2a(N-1)(2b'+m)+2b'(b'+m+1)+m+\frac23(N-1)(2N-1)a^2\Big)\,.
$$

Let us first consider the two-particle case ($N=2$), for which
the spin space $\fS$ is spanned by the four spin states
$\ket\pm\pm\equiv\ket{\pm\frac12\,}{\pm\frac12}$.
For $m=1$ the polynomial module $\BcR_1$ is the one-dimensional space
$\myspan\{\vp_1\}$, with
\begin{equation}\label{N2vp1}
  \vp_1=(z_1-z_2)\big(\ket++-\ket--\big)+(z_1+z_2)\big(\ket-+-\ket+-\big)\,.
\end{equation}
Therefore, the spin state
\begin{equation}\label{N2psi1}
\psi_1(\bx)=\frac{\mu(\bx)}{1-k^2\sn^2 x_{12}^-\sn^2 x_{12}^+}\,\Big[
\sn x_{12}^-\cn x_{12}^+\dn x_{12}^+\big(\ket++-\ket--\big)
+\sn x_{12}^+\cn x_{12}^-\dn x_{12}^-\big(\ket-+-\ket+-\big)\Big]
\end{equation}
is an eigenfunction of the Hamiltonian of type 6a,
where the gauge factor $\mu(\bx)$ is given in Eq.~\eqref{mu6a}. The
corresponding eigenvalue is $E_1=E_0-c_0$.

If $m=2$, the antisymmetrised polynomial module $\BcR_2$
is the three-dimensional space $\myspan\{\vp_1,\vp_2,\vp_3\}$,
where $\vp_1$ is given by Eq.~\eqref{N2vp1} and
\begin{equation}\label{N2vp23}
\begin{aligned}
  & \vp_2=(z_1^2-z_2^2)\big(\ket+++\ket---\ket+--\ket-+\big)\,,\\
  & \vp_3=z_1z_2(z_1-z_2)\big(\ket++-\ket--\big)
          +z_1z_2(z_1+z_2)\big(\ket+--\ket-+\big)\,.
\end{aligned}
\end{equation}
The matrix of the gauge spin Hamiltonian $\BH$ (or $H^\ast$)
in the basis $\{\vp_1,\vp_2,\vp_3\}$ is given by
$$
\begin{pmatrix}
E_0-c_0-4(1+k^2) & 0 &  8(2b+1)\\[1mm]
0 & E_0-2c_0 & 0\\[1mm]
8(2b'+1)k^2 & 0 & E_0-3c_0-4(1+k^2)
\end{pmatrix}\,,
$$
whose eigenvalues are
$$
E_{1,3}=E_0-2c_0-4(1+k^2)\mp\De\,,\qquad
E_2=E_0-2c_0\,,
$$
where $\De=\big[c_0^2+64k^2(2b+1)(2b'+1)\big]^{1/2}$. The corresponding physical
wavefunctions are
\begin{align}
&\psi_{1,3}(\bx)=\frac{\mu(\bx)}{1-k^2\sn^2 x_{12}^-\sn^2 x_{12}^+}\,
\bigg[\sn x_{12}^-\cn x_{12}^+\dn x_{12}^+\Big(c_0\mp\De
+8k^2(2b'+1)\,\sn (2x_1)\sn (2x_2)\Big)\big(\ket++-\ket--\big)\notag\\
&\qquad\qquad\:
+\sn x_{12}^+\cn x_{12}^-\dn x_{12}^-\Big(c_0\mp\De
-8k^2(2b'+1)\,\sn (2x_1)\sn (2x_2)\Big)\big(\ket-+-\ket+-\big)\bigg]
\,,\label{N2psi23}\\
&\psi_2(\bx)=\mu(\bx)\,\frac{\sn x_{12}^-\cn x_{12}^-\dn x_{12}^-
\sn x_{12}^+\cn x_{12}^+\dn x_{12}^+}
{\big(1-k^2\sn^2 x_{12}^-\sn^2 x_{12}^+\big)^2}\,
\big(\ket+++\ket---\ket+--\ket-+\big)\,.\notag
\end{align}

For $m=3$ the antisymmetrised polynomial module $\BcR_3$ is spanned by
the spin functions $\vp_1,\dots,\vp_6$, where
\begin{equation}\label{N2vp456}
\begin{aligned}
  & \vp_4=(z_1^3-z_2^3)\big(\ket++-\ket--\big)
           +(z_1^3+z_2^3)\big(\ket-+-\ket+-\big)\,,\\
  & \vp_5=z_1z_2(z_1^2-z_2^2)\big(\ket+++\ket--+\ket+-+\ket-+\big)\,,\\
  & \vp_6=z_1^2z_2^2(z_1-z_2)\big(\ket++-\ket--\big)
          +z_1^2z_2^2(z_1+z_2)\big(\ket-+-\ket+-\big)\,.
\end{aligned}
\end{equation}
The matrix representing $\BH-E_0$ in the basis $\{\vp_1,\dots,\vp_6\}$
is
$$
\begin{pmatrix}
-c_0-8(1+k^2) & 0 & 8(2b+1) & -8(4a+2b+3) & 0 & 0\\[1mm]
0 & -2c_0-8(1+k^2) & 0 & 0 & 0 & 0\\[1mm]
8(2b'+3)k^2 & 0 & -3c_0-16(1+k^2) & -16a(1+k^2) & 0 & 8(2b+3)\\[1mm]
-8(2b'+1)k^2 & 0 & 0 & -3c_0+16a(1+k^2) & 0 & -8(2b+1)\\[1mm]
0 & 0 & 0 & 0 & -4c_0-8(1+k^2) & 0\\[1mm]
0 & 0 & 8(2b'+1)k^2 & -8(4a+2b'+3)k^2 & 0 & -5c_0-8(1+k^2)
\end{pmatrix}.
$$
Clearly, $E_0-2c_0-8(1+k^2)$ and $E_0-4c_0-8(1+k^2)$ belong to the
spectrum of $\BH$ and hence of $H$, with corresponding
eigenfunctions given respectively by $\psi_2(\bx)$ in Eq.~\eqref{N2psi23} and
\begin{equation}\label{N2psi5}
\psi_5(\bx)=\mu(\bx)\,\frac{\sn x_{12}^-\cn x_{12}^-\dn x_{12}^-
\sn x_{12}^+\cn x_{12}^+\dn x_{12}^+}
{\big(1-k^2\sn^2 x_{12}^-\sn^2 x_{12}^+\big)^2}\,
\sn (2x_1)\sn (2x_2)\,
\big(\ket+++\ket--+\ket+-+\ket-+\big)\,.
\end{equation}
The remaining algebraic levels are the roots
of a fourth degree polynomial, whose expression is too long to
display here. For instance, if $a=b=b'=1$, $k^2=1/2$, and $c_0=-14$
(so that $\be=0$), the algebraic levels are approximately
$E_1=-327.4$, $E_2=-288$, $E_3=-281.4$, $E_4=-262.2$, $E_5=-260$,
and $E_6=-201.0$.\\

In the three-particle case, the spin space $\fS$ is spanned by the eight
states $\kett\pm\pm\pm$. If $m=1$, the antisymmetrised space $\BcR_1$
is trivial. For $m=2$, the antisymmetrised space $\BcR_2$ is spanned
by the single state
\begin{multline}\label{N3vp1}
\vp_1 = z_{12}^-\,z_{13}^-\,z_{23}^-\,\big(\kett++++\kett---\big)
-z_{12}^-\,z_{13}^+\,z_{23}^+\,\big(\kett++-+\kett--+\big)\\
-z_{12}^+\,z_{13}^+\,z_{23}^-\,\big(\kett+--+\kett-++\big)
+z_{12}^+\,z_{13}^-\,z_{23}^+\,\big(\kett+-++\kett-+-\big)\,,
\end{multline}
where $z_{ij}^\pm=z_i\pm z_j$. Consequently, $\psi_1(\bx)=\mu(\bx)\vp_1(\bz)$,
with $z_i=\sn (2x_i)$, is an eigenfunction of $H$ with eigenvalue
$E_0-3c_0-4(1+k^2)$.

When $m=3$, a basis for $\BcR_3$ is given by the function
$\vp_1$ in Eq.~\eqref{N3vp1} and
\begin{equation}
\begin{aligned}
& \vp_2=z_{12}^-\,z_{13}^-\,z_{23}^-\,(z_1+z_2+z_3)\,\big(\kett+++-\kett---\big)
+z_{12}^-\,z_{13}^+\,z_{23}^+\,(z_1+z_2-z_3)\,\big(\kett--+-\kett++-\big)\\
&\qquad\:
+z_{12}^+\,z_{13}^+\,z_{23}^-\,(-z_1+z_2+z_3)\,\big(\kett+---\kett-++\big)
+z_{12}^+\,z_{13}^-\,z_{23}^+\,(z_1-z_2+z_3)\,
\big(\kett+-+-\kett-+-\big),\\[1mm]
& \vp_3=z_{12}^-\,z_{13}^-\,z_{23}^-\,(z_1z_2+z_1z_3+z_2z_3)\,
\big(\kett++++\kett---\big)
+z_{12}^-\,z_{13}^+\,z_{23}^+\,(-z_1z_2+z_1z_3+z_2z_3)\,
\big(\kett--++\kett++-\big)\\
&\qquad\:
+z_{12}^+\,z_{13}^+\,z_{23}^-\,(z_1z_2+z_1z_3-z_2z_3)\,
\big(\kett+--+\kett-++\big)
-z_{12}^+\,z_{13}^-\,z_{23}^+\,(z_1z_2-z_1z_3+z_2z_3)\,
\big(\kett+-++\kett-+-\big),\\[1mm]
& \vp_4=z_1z_2z_3\,\big[z_{12}^-\,z_{13}^-\,z_{23}^-\,
\big(\kett+++-\kett---\big)
+z_{12}^-\,z_{13}^+\,z_{23}^+\,\big(\kett++--\kett--+\big)\\
&\qquad\:
+z_{12}^+\,z_{13}^+\,z_{23}^-\,\big(\kett-++-\kett+--\big)
+z_{12}^+\,z_{13}^-\,z_{23}^+\,\big(\kett-+--\kett+-+\big)\big].
\end{aligned}
\end{equation}
The matrix of $\BH$ in the basis $\{\vp_1,\dots,\vp_4\}$ reads
\[
\begin{pmatrix}
E_0-3c_0-16(1+k^2) & 0 & 8(4a+2b+3) & 0\\[1mm]
0 & E_0-4c_0+8(1+k^2)(2a-1) & 0 & 8(2b+1)\\[1mm]
8(2b'+1)k^2 & 0 & E_0-5c_0+8(1+k^2)(2a-1) & 0\\[1mm]
0 & 8(4a+2b'+3)k^2 & 0 & E_0-6c_0-16(1+k^2)
\end{pmatrix}\,.
\]
The eigenvalues of this matrix are
\[
E_{1,3}=E_0-4c_0+4(1+k^2)(2a-3)\mp\De_-\,,\qquad
E_{2,4}=E_0-5c_0+4(1+k^2)(2a-3)\mp\De_+\,,
\]
where
\begin{align*}
\De_+&=\Big[\big(c_0+4(1+k^2)(2a+1)\big)^2+64k^2(2b+1)(4a+2b'+3)\Big]^{1/2},
\\
\De_-&=\Big[\big(c_0-4(1+k^2)(2a+1)\big)^2+64k^2(2b'+1)(4a+2b+3)\Big]^{1/2}\,.
\end{align*}
The corresponding eigenfunctions are
\begin{align*}
& \psi_{1,3}(\bx)=\mu(\bx)
\Big[\Big(c_0-4(1+k^2)(2a+1)\mp\De_-\Big)\,\vp_1(\bz)
+8k^2(2b'+1)\,\vp_3(\bz)\Big]\,,\\[1mm]
& \psi_{2,4}(\bx)=\mu(\bx)
\Big[\Big(c_0+4(1+k^2)(2a+1)\mp\De_+\Big)\,\vp_2(\bz)
+8k^2(4a+2b'+3)\,\vp_4(\bz)\Big]\,,
\end{align*}
with $z_i=\sn (2x_i)$.

\section*{Appendix}

In this Appendix we present a list of identities satisfied
by the Dunkl operators~\eqref{J-}--\eqref{J+} used to compute
the gauge spin Hamiltonian~\eqref{BH}.
\begin{align*}
& \sum_i (J_i^-)^2=\sum_i \pa_{z_i}^2
+4a\sum_{i\neq j}\frac{z_i}{z_i^2-z_j^2}\,\pa_{z_i}
+2b\sum_i \frac1{z_i}\,\pa_{z_i}
+a\sum_{i\neq j}\frac{K_{ij}-1}{(z_i-z_j)^2}
+a\sum_{i\neq j}\frac{\tK_{ij}-1}{(z_i+z_j)^2}
+b\sum_i\frac{K_i-1}{z_i^2},\\
& \sum_i (J_i^0)^2=\sum_i\Big(\pa_{z_i}^2
+\big(1-m+2a(1-N)\big)z_i\pa_{z_i}\Big)
+4a\sum_{i\neq j}\frac{z_i^3}{z_i^2-z_j^2}\,\pa_{z_i}
+a\sum_{i\neq j}\,\frac{z_iz_j}{(z_i-z_j)^2}\,(K_{ij}-1)\\
&\qquad\quad -a\sum_{i\neq j}\,\frac{z_iz_j}{(z_i+z_j)^2}\,(\tK_{ij}-1)
+\frac{Nm^2}4+\frac{a^2}{12}
\Bigg({\sum_{i,j,k}}'\big[4-(K_{ij}+\tK_{ij})(K_{ik}+\tK_{ik})\big]
+6\sum_{i\neq j}(1-K_iK_j)\Bigg),\\
& \sum_i (J_i^+)^2=\sum_i\Big(\pa_{z_i}^4
-2\big(b'+m-1+2a(N-1)\big)z_i^3\pa_{z_i}\Big)
+4a\sum_{i\neq j}\frac{z_i^5}{z_i^2-z_j^2}\,\pa_{z_i}
+a\sum_{i\neq j}\,\frac{z_i^2z_j^2}{(z_i-z_j)^2}\,(K_{ij}-1)\\
&\qquad\quad +a\sum_{i\neq j}\,\frac{z_i^2z_j^2}{(z_i+z_j)^2}\,(\tK_{ij}-1)
+b'\sum_i z_i^2\big((-1)^m K_i-1\big)+m(m-1+2b')\sum_i z_i^2\,,\\
&\sum_i J_i^0=\sum_i z_i\pa_{z_i}-\frac{Nm}2\,.
\end{align*}
The symbol $\sum_{i,j,k}'$ means summation in $i,j,k$ with
$i\neq j\neq k\neq i$.

\begin{acknowledgments}
This work was partially supported by the DGES under grant PB98-0821.
R. Zhdanov would like to acknowledge the financial support
of the Spanish Ministry of Education and Culture during his
stay at the Universidad Complutense de Madrid.
\end{acknowledgments}

\end{document}